\documentclass{jpsj-suppl}
\usepackage{txfonts} 
\usepackage{bm}
\newcommand {\beq}{\begin{equation}}
\newcommand {\eeq}{\end{equation}}
\newcommand {\bea}{\begin{eqnarray}}
\newcommand {\eea}{\end{eqnarray}}
\newcommand {\nn}{\nonumber \\}
\newcommand {\e}{{\rm e}}

\newcommand {\m}{\mu}
\newcommand {\n}{\nu}
\newcommand {\pl}{\partial}
\newcommand {\p} {\phi}
\newcommand {\vp}{\varphi}

\newcommand {\al}{\alpha}

\newcommand {\Ga}{\Gamma}

\newcommand {\ka}{\kappa}
\newcommand {\la}{\lambda}
\newcommand {\La}{\Lambda}
\newcommand {\si}{\sigma}

\newcommand {\om}{\omega}

\newcommand {\ep}{\epsilon}

\newcommand {\na}{\nabla}
\newcommand {\del}  {\delta}

\newcommand {\mn}{{\mu\nu}}

\newcommand {\half}{ {\frac{1}{2}} }

\newcommand {\sqg} {\sqrt{-g}}

\newcommand {\Lcal}{{\cal L}}

\newcommand {\Dcal}{{\cal D}}

\newcommand {\fvec}{{\vec f}}

\newcommand {\vvec}{{\vec v}}

\newcommand {\pvec}{{\vec p}}

\newcommand {\xvec}{{\vec x}}

\newcommand {\navec}{{\vec \nabla}}

\newcommand {\Dtil}{{\tilde D}}

\newcommand {\gtil}{{\tilde g}}

\newcommand {\ttil} {{\tilde t}}

\newcommand {\Ptil}{{\tilde \Phi}}

\newcommand {\natil} {{\tilde \nabla}}

\newcommand {\Lhat}{{\hat L}}

\newcommand {\delh} {{\hat \delta}}

\newcommand {\rdot}{\dot{r}}

\newcommand {\xdot}{\dot{x}}

\newcommand {\intfx} {{\int d^4x}}

\newcommand {\change} {\leftrightarrow}
\newcommand {\ra} {\rightarrow}

\newcommand {\com}  {{\quad ,}}
\newcommand {\q}    {\quad}

\newcommand {\nl}    {\newline}

\newcommand {\Phicl}  {\Phi_{cl}}

\newcommand {\gmn} {g_\mn}
\newcommand {\Lcalmat} {\Lcal_{mat}}
\newcommand {\HU} {H_0}

\title{New Approach to Cosmological Fluctuation using the Background Field Method and 
CMB Power Spectrum}

\author{Shoichi \textsc{Ichinose}}

\inst{Laboratory of Physics, School of Food and Nutritional Sciences, 
University of Shizuoka\\
Yada 52-1, Shizuoka 422-8526, Japan
}

\email{ichinose@u-shizuoka-ken.ac.jp}

\recdate{July 15, 2013}

\abst{
A new field theory formulation is presented for the analysis of 
the CMB power spectrum distribution in the cosmology. 
The background-field formalism is fully used. 
Stimulated by the recent idea of the {\it emergent} gravity,  
the gravitational (metric) field $g_\mn$ is 
{\it not} taken as the quantum-field, but 
 as the background field. 
The statistical fluctuation effect of the metric field is taken into 
account by the path (hyper-surface)-integral over the space-time. 
Using a simple scalar model on the curved (dS$_4$) space-time, 
we explain the above things with the following additional points: \ 
1) Clear separate treatment of the classical effect, the statistical effect 
and the quantum effect; \ 
2) The cosmological fluctuation comes {\it not} from the 'quantum' gravity but 
from the unkown 'microscopic' movement; \ 
3) IR parameter ($\ell$) is introduced for the time axis as the periodicity. 
Time reversal(Z$_2$)-symmetry is introduced in order to treat the problem separately 
with respect to the Z$_2$ parity. This procedure much helps both UV and IR regularization to work well. 
}

\kword{
CMB power spectrum, statistical fluctuation, background-field method, 
Bunch-Davies vaccum, path-integral
}

\begin{document}
\maketitle

\section{Introduction\label{intro}}
The more the data of the CMB experiment accumulates, 
the richer structure of our universe is exposed. 
In this fascinating period, 
the CMB spectrum data plays so important role to understand our universe. 
We approach this spectrum using the background-field method[1] 
which keeps the important position in the quantum field theory. 

The wonderful development of the string theory and D-brane theory has brought us 
the intimate relation between gravity and (condensed-)matter physics through 
AdS/CFT correspondence. Especially the relation to the hydrodynamics gradually 
becomes important through the discussion about the ratio of viscosity and entropy. 
As the continuum field theory, 
the gravitational theory and the fluid theory share common problems such 
as the velocity distribution in the galaxy and that in the rotated (viscous) fluid.  
Another important recent trend is about the view to the gravitational (metric) field. 
The standpoint "emergent gravity"[2] is gradually taken seriously. It claims 
gravity is {\it not} fundamental one, but emerges from the statistical property, such 
as entropy, of the medium (entropic force). 

With these new trends of the gravitational theory, we newly formulate the power spectrum 
calculation. The spectrum tells us how the light coming from one direction correlates 
with the other one that comes from another direction. Generally the correlation between 
the lights from $n$ directions is described by certain n-point function. It corresponds to S-matrix 
in the quantum field theory. We formulate the $n$-point function using the background-field 
formalism[1].


\section{CMB power spectrum in the background-field formalism\label{CMB}}
Let us consider the coupled system of the scalar field $\Phi(x)$ and the gravitational (metric) field 
$g_\mn(x)$ in the 3(space)+1(time) dimensional manifold. 
\ \ 
$\ \bm{<<}\ 
S[\Phi;g_\mn]
=\intfx\sqg (
{-1}/{16\pi G_N}\cdot \\ 
(R-2\la)+\Lcalmat), 
\Lcalmat\equiv -1/2\cdot\na_\m\Phi\na^\m\Phi-{m^2}/{2}\cdot\Phi^2-V(\Phi) 
,\  
V(\Phi)={\si}/{4!}\cdot\Phi^4
,
   \ (1)\bm{>>}\ $
\ \ 
where  $\si>0$, 
 $G_N$ is the Newton's gravitational constant, $m$ is the scalar mass, and $\la$ is 
the cosmological constant.
$V(\Phi)$ is the scalar potential. $\Lcalmat$ is the matter($\Phi$) part 
of the Lagrangian. The used notaion is\  
$
(x^\m)=(x^0, x^1, x^2, x^3)\equiv (t,x,y,z),\  -\infty < t, x, y, z < +\infty
$. 
We expand the scalar-field around its {\it classical background} field ($\Phicl$)
in order to {\it quantize} the matter-field in the background-field 
formalism:\ 
$ \Phi=\Phicl+\vp $,\ 
where $\Phicl$ is the classsical solution of the matter part ((B) of (3)). 
The {\it effective action} $\Ga[\Phicl;\gmn]$ is defined as 
\ \ 
$\ \bm{<<}\ 
\e^{i\Ga[\Phicl;\gmn]}=\int\Dcal\vp\exp i\left\{ 
S[\Phicl+\vp;\gmn]-{\del S[\Phicl;\gmn]}/{\del\Phicl}\cdot\vp 
                                        \right\} 
.
\ (2)\bm{>>}\ $
\ \ 
Here we do {\it not} Taylor-expand 
the gravitational field, which means, in the present treatment, 
the gravitational field $g_\mn(x)$ is {\it not field-quantized} and is treated as 
the background (classical) field.
$\vp$ is the {\it quantum field}. The scalar (matter) field only is quantized. 

$\Phicl$ and $\gmn$ must satisfy the {\it on-shell condition}:
\ \ 
$\ \bm{<<}\ 
\mbox{On-shell Condition}:\ 
\mbox{(A)}\ \ {\del \Ga[\Phicl;\gmn]}/{\del\Phicl}=0;\ 
\mbox{(B)}\ \ {\del \Ga[\Phicl;\gmn]}/{\del\gmn}=0
.
\ (3)\bm{>>}\ $
\ \ 
In the present work, $\Phicl$ and $\gmn$ are the {\it background fields}. 
At the {\it tree level} (of the scalar quantum-loop expansion), Eq.(A) is 
\ \ 
$\ \bm{<<}\ 
{\del S[\Phicl;\gmn]}/{\del\Phicl}=0; 
\sqg \{\na^2\Phicl 
-m^2\Phicl\} =\sqg {\del V(\Phicl)}/{\del\Phicl} 
.
\ (4)\bm{>>}\ $
\ \ 
The above solution $\Phicl$ is formally obtained as
\ \ 
$\ \bm{<<}\ 
\Phicl(x)=\Phi_0(x)+\int D(x-x') \sqg ({\del V(\Phicl)}/{\del\Phicl})|_{x'} d^4x',\ 
\sqg (\na^2-m^2)D(x-x')=\del^4(x-x')
,
\ (5)\bm{>>}\ $
\ \ 
where $\Phi_0$ is the {\it free} field, $\sqg (\na^2-m^2)\Phi_0=0$,\ and $D(x-x')$ is the {\it propagator} on the curved geometry $\gmn$. 
(
$\Phi_0$ is later used as the external source to generate the n-point function of the spectrum.
)\ 
The above equation gives the {\it tree graph} expansion where the expansion parameter 
is a small coupling ($\sigma$ in (1))  
in $V(\Phicl)$. (c.f. Loop-expansion (2) is that of the power of $\hbar$. )

On the other hand, Eq.(B) of (3) is, at the {\it tree level}, written as 
\ \ 
$\ \bm{<<}\ 
{\del S[\Phicl;\gmn]}/{\del\gmn}=0;\ 
R_\mn-1/2\cdot R\gmn+\la g_\mn=8\pi G_N T_\mn,\ 
T^\mn={2}/{\sqg}\cdot {\del(\sqg\Lcalmat)}/{\del\gmn},\ 
T_\mn=\pl_\m\Phicl\pl_\n\Phicl-\gmn \left(
1/2\cdot\pl^\si\Phicl\pl_\si\Phicl+{m^2}/{2}\cdot\Phicl^2+V(\Phicl)
                                     \right)
.
\ (6)\bm{>>}\ $

\section{dS$_4$ Geometry, Bunch-Davies Vacuum and Casimir Energy\label{BDvac}}
\subsection{dS$_4$ Geometry and Conformal Time}
 The metric $\gmn$ must satisfy the on-shell condition (3), or 
(4) and (6) at the tree level. 
Now we restrict the form of $g_\mn$ from the requirement of the space-time symmetry: 
{\it homogenity} and {\it isotropy}. 
$\ \bm{<<}\ 
ds^2=g_\mn(x)dx^\m dx^\n=-dt^2+a(t)^2(dx^2+dy^2+dz^2),\ a(t)=\e^{\rho (t)}
,
\ (7)\bm{>>}\ $
\ \ 
where we take the {\it spacially flat} case.
If we take $a(t)$ as follows :
\ \ 
$\ \bm{<<}\ 
{\rho(t)}/{t}= \HU (\mbox{constant})>0\com\q \e^{\HU t}\equiv a_{infl}(t)\com\q
\left( g^{infl}_\mn\right)=\mbox{diag}
(-1, \e^{2\HU t}, \e^{2\HU t}, \e^{2\HU t}) 
\com\q
ds^2=g_\mn^{infl}(x)dx^\m dx^\n=-dt^2+\e^{2\HU t}(dx^2+dy^2+dz^2)\com\q
-\infty < t,x,y,z < \infty\com
\ (8)\bm{>>}\ $
\ \ 
the space-time geometry becomes 4dim de Sitter (dS$_4$) which describes the {\it inflation universe}. 
$H_0$ is a positive constant (Hubble constant). 

In the following of this section, we consider the dS$_4$ metric, $g_\mn^{infl}$, as the background field. 
We now transform from 
$t$ to the {\it conformal time} $\eta$ defined by 
\ \ 
$\ \bm{<<}\ 
d\eta =\e^{-\HU t}dt\com\q \eta=-{1}/{\HU}\cdot\e^{-\HU t}\com\q -\infty < t < \infty\com\q -\infty < \eta < 0
.
\ (9)\bm{>>}\ $
\ \ 
The metric transforms to the conformally-flat type. 
\ \ 
$\ \bm{<<}\ 
ds^2=g_\mn^{infl}(x)dx^\m dx^\n
=\gtil_\mn(\chi)d\chi^\m d\chi^\n ={1}/{(\HU\eta)^2}\cdot (-d\eta^2+dx^2+dy^2+dz^2),\ 
\sqrt{-\gtil}={1}/{(\HU\eta)^4},\ 
(\chi^\m)=(\chi^0,\chi^1,\chi^2,\chi^3)=(\eta,x,y,z),
\ (10)\bm{>>}\ $
\ \ 
The perturbative solution $\Phicl$, (5), is given by 
\ \ 
$\ \bm{<<}\ 
\Phicl(\chi)=\Phi_0(\chi)+\int \Dtil(\chi,\chi')~\left.\frac{1}{(\HU \eta')^4}\frac{\del V(\Phicl)}{\del\Phicl}\right|_{\chi'} d^4\chi',\ 
\sqrt{-\gtil} (\natil^2-m^2)\Phi_0=
-\left\{\pl_\eta\frac{1}{(\HU\eta)^2}\pl_\eta+\frac{m^2}{(\HU\eta)^4}
-\frac{1}{(\HU\eta)^2}\navec^2   \right\}\Phi_0=0
,
\ (11)\bm{>>}\ $
\ \ 
where $\natil^2=\gtil^\mn \natil_\m\natil_\n \ (\natil_\m=\pl/\pl\chi^\m+\cdots)$ and 
$\Phi_0$ is the {\it free} field.  $\Dtil(\chi,\chi')$ is the {\it propagator} on 
the dS$_4$ geometry $\gtil_\mn(\chi)$ ( or $g^{infl}_\mn(x)$).
\ \ 
$\ \bm{<<}\ 
\sqrt{-\gtil} (\natil^2-m^2)\Dtil(\chi,\chi')= \\ 
-\left\{\pl_\eta\frac{1}{(\HU\eta)^2}\pl_\eta+\frac{m^2}{(\HU\eta)^4}
-\frac{1}{(\HU\eta)^2}\navec^2   \right\}\Dtil(\chi,\chi')=
\del^4(\chi-\chi')
,
\ (12)\bm{>>}\ $
\ \

\subsection{Z$_2$ Symmetry, Periodicity and IR parameter $\ell$}
In the (matter-field) quantization on dS$_n$ (and AdS$_n$) geometry, the control of 
the IR-divergence is important[3]. 
In order to {\it regularize} the IR behavior, we introduce 
the following symmetries in the time coordinate $t$: 
\ \ 
$\ \bm{<<}\ 
\mbox{Z}_2\mbox{ Symmetry}:\ \ t\ \change\ -t,\ \ 
\mbox{Periodicity}:\ \ t\ra t+2\ell 
,
\ (12b 
)\bm{>>}\ $
\ \ 
where $\ell$ is the {\it period} parameter (IR parameter).
As for the conformal
time, we {\it redefine} it as follows, 
\ \ 
$\ \bm{<<}\ 
\eta=\left\{
\begin{array}{cccc}
-{1}/{H_0}\cdot\e^{-H_0t}\ , & d\eta=-H_0\eta dt\ , &  0 < t < \ell\ , 
                                                       &{-1}/{H_0}\cdot <\eta<{-1}/{\om} \\
+{1}/{H_0}\cdot\e^{H_0t}\ ,  & d\eta=H_0\eta dt\ ,  &  -\ell < t < 0\ , 
                                                       & {1}/{\om}<\eta<{1}/{H_0}
\end{array}
\right.
\ (13)\bm{>>}\ $
\ \ 
($
\om\equiv \e^{\ell H_0}H_0\gg\ H_0
$)
which leads to the relation $\eta\change -\eta$ corresponding to the Z$_2$ symmetry 
in (12b). 
In this definition the far past ($t\ra -\ell$) corresponds to $\eta\ra 1/\om$, the far 
future ($t\ra +\ell$) corresponds to $\eta\ra -1/\om$. 
$t=\mp 0$ correspond to the singular points $\eta=\pm 1/\HU$. 
See Fig.1. 
\begin{figure}[h]
\begin{minipage}{10pc}
\includegraphics[width=10pc,height=10pc]{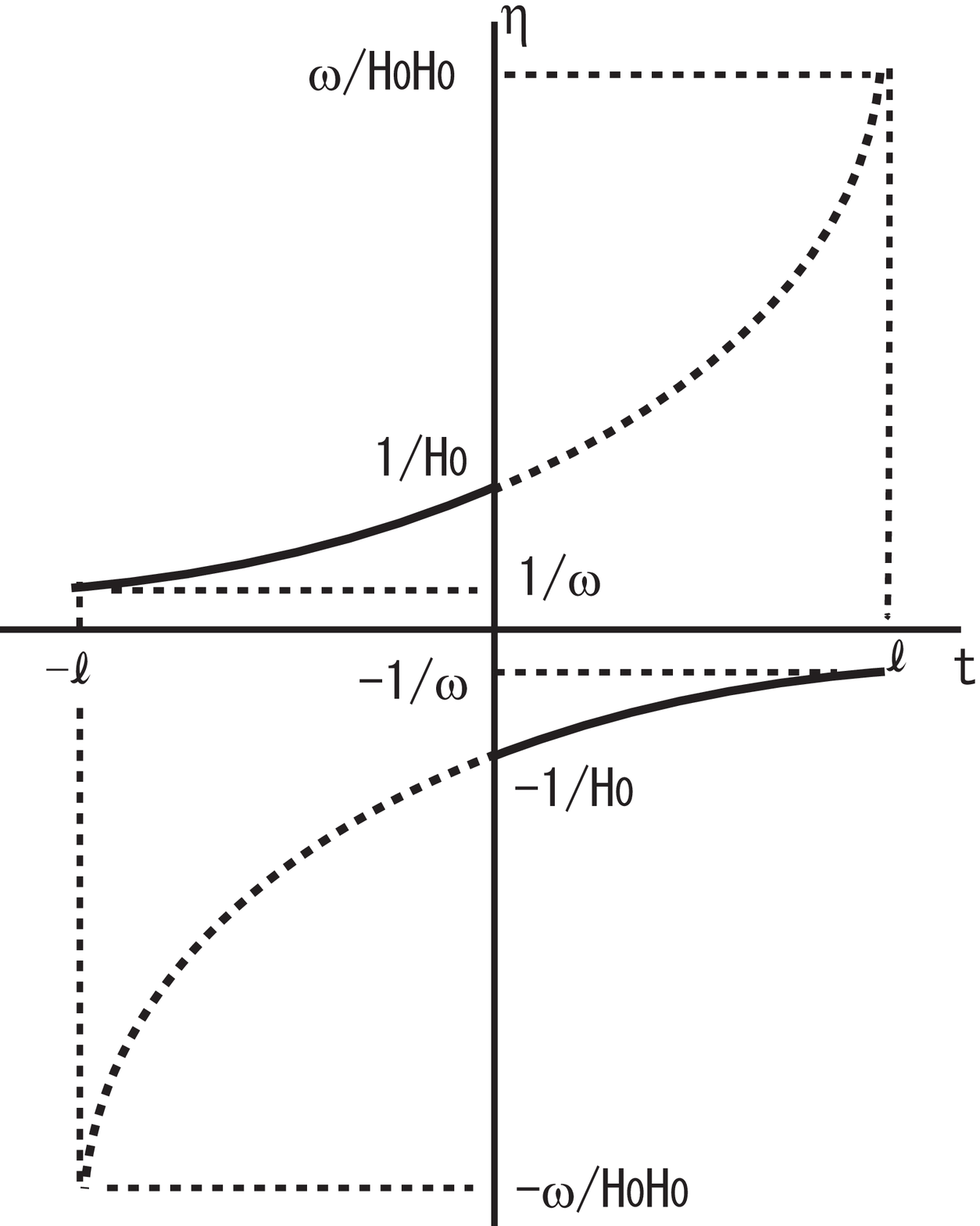}
\caption{\label{ConfTime2}
Conformal time ($\eta$) versus ordinary time ($t$). Time-reversal 
correspondence is valid. Eq.(13)
}
\end{minipage}
\hspace{1pc}
\begin{minipage}{10pc}
\includegraphics[width=10pc]{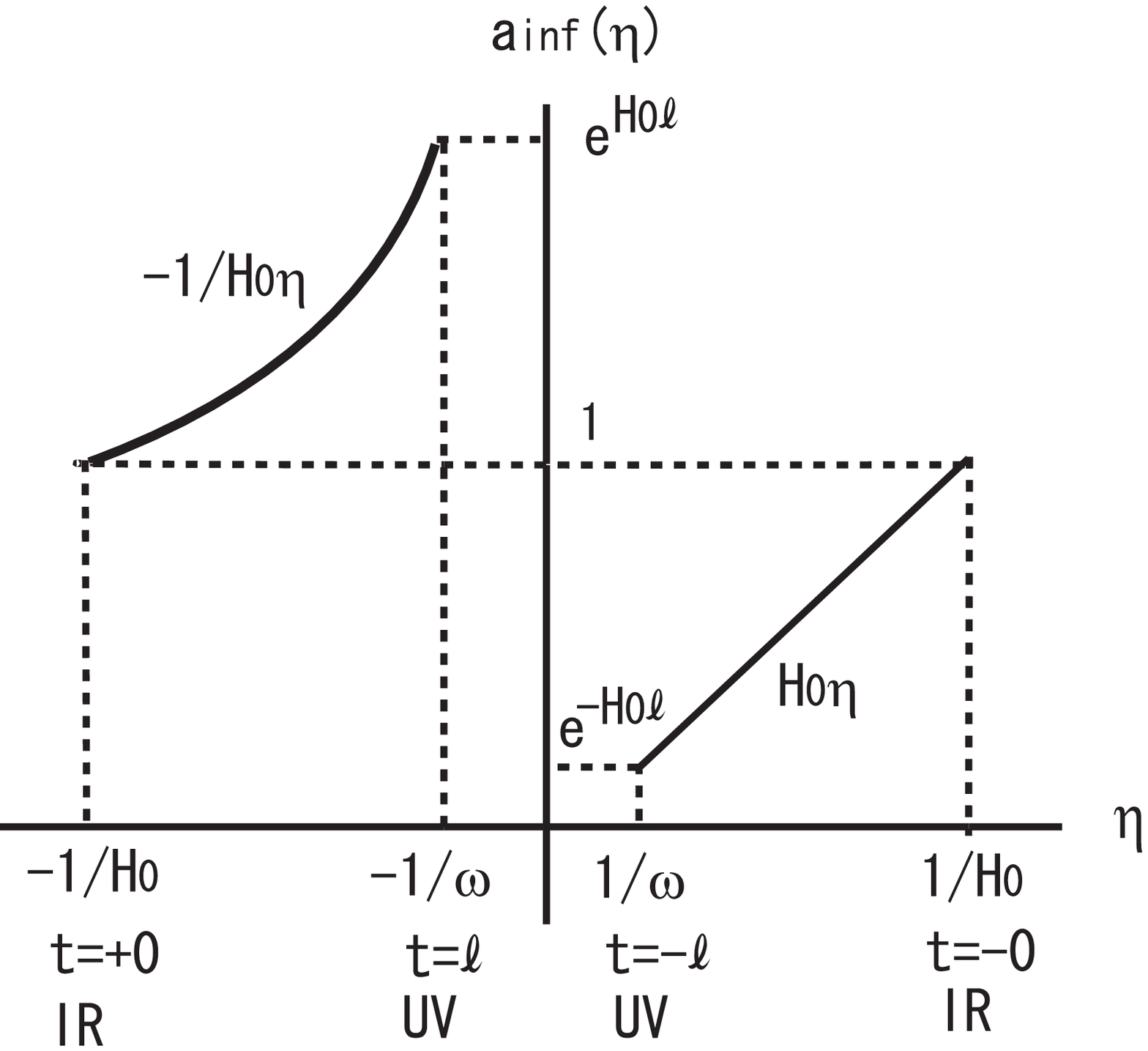}
\caption{\label{aINF}
Scale factor in dS$_4$ geometry. Eq.(14).
}
\end{minipage} 
\hspace{1pc}
\begin{minipage}{10pc}
\includegraphics[width=10pc]{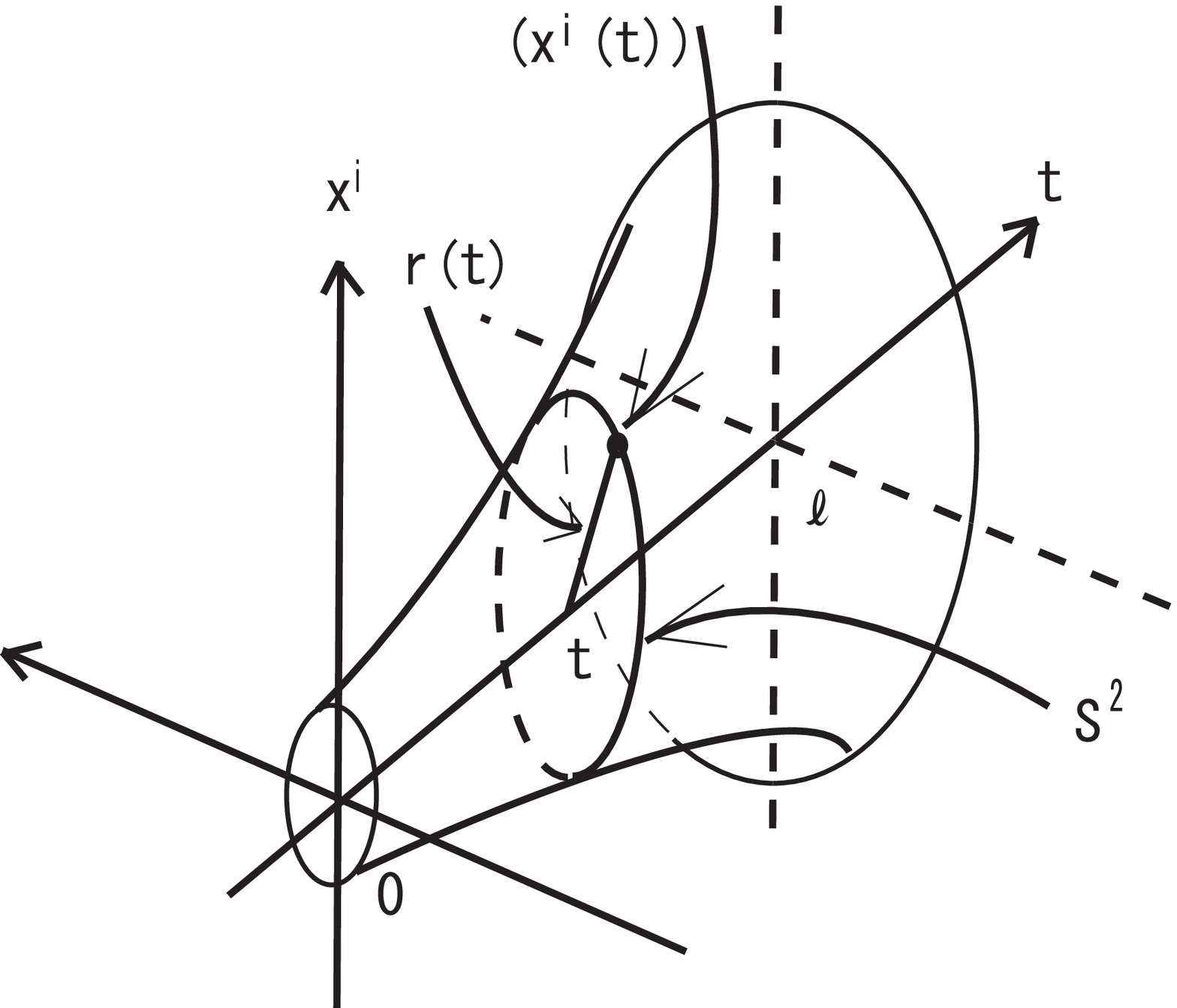}
\caption{\label{HyperSurf}
Hyper-surface in dS$_4$ space-time. Eq.(23b).
}
\end{minipage}
\end{figure}
Hence $a_{inf}(t)=\e^{H_0t}$ is expressed as \nl 
$\ \bm{<<}\ 
0<t<\ell\ (-{1}/{H_0}<\eta<{-1}/{\om})\ :\q
\eta=-{\e^{-H_0t}}/{H_0}\com\q d\eta=-H_0\eta dt\com\q a_{inf}(\eta)=-{1}/{H_0\eta}\nn
-\ell<t<0\ ({1}/{\om}<\eta<{1}/{H_0})\ :\q 
\eta={\e^{H_0t}}/{H_0}\com\q d\eta=H_0\eta dt\com\q a_{inf}(\eta)=H_0\eta
\ (14)\bm{>>}\ $
\ \ 
See Fig.2. 

Let us switch, from $\Phi_0(\eta,\xvec)$ and $\Dtil(\chi,\chi')$, to the spacially-Fourier-transformed expression
$\phi_{\pvec}(\eta)$ and $\Dtil_\pvec(\eta,\eta')$:
\ \ 
$\ \bm{<<}\ 
\Phi_0(\eta,\xvec)=
\int{d^3\pvec}/{(2\pi)^3}\cdot\e^{i\pvec\cdot\xvec}\phi_{\pvec}(\eta),\ 
\Dtil(\chi,\chi')=
\int{d^3\pvec}/{(2\pi)^3}\cdot\e^{i\pvec\cdot(\xvec-\xvec')}\Dtil_\pvec(\eta,\eta')
,
\ (15)\bm{>>}\ $
\ \ 
where $\Dtil_\pvec(\eta,\eta')$ is called 
'Momentum/Position propagator'[3]. 
From (11), $\phi_{\pvec}(\eta)$ satisfies the following {\it Bessel} 
eigenvalue equation. 
\ \ 
$\ \bm{<<}\ 
\left\{
{\pl_\eta}^2-{2}/{\eta}\cdot\pl_\eta+{m^2}/{(\HU\eta)^2}
+M^2   
\right\}\p_M(\eta)
=
\{ s(\eta)^{-1}\Lhat_\eta + M^2\}\phi_M(\eta)=0,\  M^2\equiv\pvec^2 ,\ 
s(\eta)\equiv {1}/{(\HU \eta)^2},
\ (16)\bm{>>}\ $
\ \ 
where 
$ \Lhat_\eta\equiv \pl_\eta s(\eta)\pl_\eta
+{m^2}/{(\HU\eta)^4}$. 
From (12), $\Dtil_\pvec(\eta,\eta')$ satisfies\nl
$\ \bm{<<}\ 
 \{ \Lhat_\eta +\pvec^2 s(\eta)\}\Dtil_{\pvec}^\mp(\eta,\eta')=
\left\{
\begin{array}{cc}
\ep(\eta)\ep(\eta')\delh(|\eta|-|\eta'|)  &  \mbox{for P=}-  \\
\delh(|\eta|-|\eta'|)   &   \mbox{for P=}+
\end{array}
\right.
\ (17)\bm{>>}\ $
\ \ 
where $\ep(\eta)$ is the sign function. 

The Bessel equation (16) gives us the free field wave function as\ \ 
$
\p_M(\eta)=\eta^{3/2}Z_\n(M\eta)\com\q \n=
\sqrt{(3/2)^2-(m/\HU)^2}
$,\ 
where $\n=0, 1/2, \sqrt{5}/2,$ and $3/2$ correspond to $m=(3/2)\HU, \sqrt{2}\HU, \HU,$ and $0$ 
respectively. When $m$ is non-negative real number, the scalar mass has 
the {\it upper-bound}:\ $0\leq m\leq (3/2)\HU$. (We may also consider the imaginary mass:\ $m^2<0$. ) 

\subsection{Boundary Condition, Bunch-Davies Vacuum and Casimir Energy}
As for the boundary condition for Z$_2$-parity {\it odd} free field ($P=-$), we 
take the following one based on the requirement of the {\it continuity} 
at $\eta=\pm 1/\om,\pm 1/\HU$ ($t=\mp \ell, \mp 0$):\ 
$
\Phi_0(\eta\ra\pm 1/\om, \xvec)=0 (\mbox{Dirichlet}),\ 
\Phi_0(\eta\ra\pm 1/\HU, \xvec)=0 (\mbox{Dirichlet})
$. 
As for the boundary condition for Z$_2$-parity {\it even} case ($P=+$), we 
take the following one based on the requirement of the {\it smoothness} 
at $\eta=\pm 1/\om,\pm 1/\HU$ ($t=\mp \ell, \mp 0$):\ 
$
\left.\pl_\eta\Phi_0\right|_{\eta\ra\pm 1/\om}=0 (\mbox{Neumann}),\ 
\left.\pl_\eta\Phi_0\right|_{\eta\ra\pm 1/\HU}=0 (\mbox{Neumann})
$.\nl

 Casimir energy is given by $\si$-independent part 
 (free part) of the 1-loop effective action  in (2). 
$\ \bm{<<}\ 
\exp [i\Ga^{\mbox{1-loop}}]
=\exp\left[
\int {d^3\pvec}/{(2\pi)^3}\cdot 2\int_{-1/\HU}^{-1/\om}d\eta
\{-1/2\cdot\ln(-s(\eta)^{-1}\Lhat_\eta-\pvec^2)\}
     \right]
=\exp \int_0^\infty {d\tau}/{\tau}\cdot 1/2\cdot \Tr H_\pvec(\eta,\eta';\tau)     
,
\ (18)\bm{>>}\ $
\ \ 
where $H_\pvec(\eta,\eta';\tau)$ is the Heat-Kernel: 
\ \ 
$\ \bm{<<}
\{ 
{\pl}/{\pl\tau}-(s^{-1}\Lhat_\eta+\pvec^2)
\}
H_\pvec(\eta,\eta';\tau)=0,\ 
H_\pvec(\eta,\eta';\tau)=(\eta|\e^{(s^{-1}\Lhat_\eta+\pvec^2)\tau}|\eta')
.
\ (19)\bm{>>}$
\ \ 
This is a formal expression using Dirac's abstract states $|\eta), (\eta|$. 
It is rigorously defined by using 
the complete and orthogonal eigen functions \{ $\phi_n(\eta)$\} of the operator 
$s^{-1}\Lhat_\eta$.  
\nl
$\ \bm{<<}\ 
\phi_n(\eta)\equiv (n|\eta)=(\eta|n),\ 
\{s(\eta)^{-1}\Lhat_\eta+{M_n}^2\}\phi_n(\eta)=0,\ 
(\eta|\eta')=\left\{
\begin{array}{lc}
(\HU\eta)^2\ep(\eta)\ep(\eta')\delh(|\eta|-|\eta'|) & \mbox{for  }P=-\\
(\HU\eta)^2\del(|\eta|-|\eta'|) & \mbox{for  }P=+\\
\end{array}
              \right.
              \nn
\left(
\int_{-1/\HU}^{-1/\om}+\int_{1/\om}^{1/\HU}
\right)
\frac{d\eta}{(\HU \eta)^2}
(n|\eta)(\eta|k)=2\int_{-1/\HU}^{-1/\om}\frac{d\eta}{(\HU \eta)^2}(n|\eta)(\eta|k)
=(n|k)=\del_{n,k}\com\nn
\left(
\int_{-1/\HU}^{-1/\om}+\int_{1/\om}^{1/\HU}
\right)
\frac{d\eta}{(\HU \eta)^2}
|\eta)(\eta|=2\int_{-1/\HU}^{-1/\om}\frac{d\eta}{(\HU \eta)^2}|\eta)(\eta|
=1\com\q
\sum |n)(n|=1              
.
\ (20)\bm{>>}\ $
\nl
The set \{ $\phi_n(\eta)$\} constitutes Bunch-Davies vacuum.

\subsection{Wick Rotation for $\pvec$}\label{wick}
From the previous result, we evaluate Casimir energy 
of the dS$_4$ space-time. 
\ \ 
$\ \bm{<<}\ 
-\HU^{-3}E_{Cas}^{dS4}
=\int {d^3\pvec}/{(2\pi)^3}\cdot 2\int_{-1/\HU}^{-1/\om}d\eta
  \{
1/2\cdot\int^\infty_0 {d\tau}/{\tau}\cdot (\eta|\e^{\tau(s(\eta)^{-1}\Lhat_\eta +\pvec^2)}|\eta)
  \}
= \int_0^\infty {d\tau}/{\tau}\cdot 1/2\cdot \Tr H_\pvec(\eta,\eta';\tau)     
.
\ (21)\bm{>>}\ $
\ \ 
This expresssion diverges very badly (UV-divergence). 
To regularize it, we do
\nl
$\ \bm{<<}\ 
\mbox{Wick rotation for {\it space}-components of momentum}\q
p_x\ ,\ p_y\ ,\ p_z\ \longrightarrow ip_x\ ,\ ip_y\ ,\ ip_z
\ (22)\bm{>>}\ $
The regularized expression is the same as Casimir energy for (E)AdS$_4$. 
The regularized one behaves milder but still diverges. 


\section{Metric Fluctuation and 
Averaging Over the 4D Space-Time using the Generalized Path-Integral 
\label{fluct}}
We note again the metric field $g_\mn(x)$ is treated as 
the background one. It is defined by the variation 
equation of $\Ga[\Phicl;\gmn]$, (B) of (3). 
Let us express the effective action  
as the integral of the space-time coordinates $x^\m$:\  
$
S = \Ga[\Phicl(x);\gmn(x)]\equiv\int d^4x\Lcal[x^\m]
$. 
We regard this quantity as an action for a {\it statistic-mechanical} system 
composed of its dynamical 
variables ($x^i$: $i=1,2,3$) and time ($x^0=t$). Here we consider 
the small fluctuation of coordinates $x^i$, {\it keeping $x^0=t$ fixed}, in the dS$_4$ 
geometry $g^{inf}_\mn (x)$:\  
$
x^i\ra x^i+\sqrt{\ep}f^i(\xvec,t)={x^i}',\  t=t' \ (x^0={x^0}')
$,\ 
where 
$\ep$ is a small positive parameter for dictating the perturbation order. 
(
We regard the present system not as the (metric) field theory but as the 
statistic-mechanical system of space coordinates ${\bf x}$ and time $t$. Hence this 
fluctuation should not be regarded as the gauge variation. 
)
This fluctuation 
can be absorbed into the {\it metric fluctuation} (around $g^{inf}_\mn$) as the requirement 
of the invariance of the line element ({\it general coordinate invariance}). 
\nl
$\ \bm{<<}\ 
g^{inf}_\mn(x)d{x^\m}'d{x^\n}'={g_\mn}'(x)dx^\m dx^\n\com\q
{g_\mn}'(x)=g_\mn^{inf}(x)+\ep h_\mn(x)
,\ 
h_{00}=\e^{2\HU t}\pl_0f^i\cdot\pl_0f^i\com\nn 
h_{0i}=h_{i0}=\e^{2\HU t}\pl_if^j\cdot\pl_0f^j,\ 
h_{ij}=\e^{2\HU t}\pl_if^k\cdot\pl_jf^k,\ 
\{\half(\pl_if^j+\pl_jf^i)dx^j+2\pl_0f^idt\}dx^i=0
.
\ (23)\bm{>>}\ $
We see the coordinates-fluctuation produces the metric-
fluctuation 
(around the homogeneous and isotropic (dS$_4$) metric)
, as far as the above constraint is preserved. 
The constraint comes from the difference in the perturbation order 
between the metric fluctuation ($\ep$) and the coordinate fluctuation ($\sqrt{\ep}$).  

A cause of the fluctuation is the effect 
of the underlying {\it unknown} 'micro' dynamics (just like Brownian motion 
of nano-particles in liquid and gas in the early days of the atomic physics). 
We treat it as the {\it statistical fluctuation phenomena}.   
We adopt the following strategy to compute the fluctuation effect. 
The coordinates are fluctuating 
(
, and the metric is also fluctuating as stated in the previous paragraph, 
)
in a statistical ensemble. 
In order to compute the statistical average, we must specify the statistical distribution. 
We note again that this effect is treated, in the present standpoint, 
not as the quantum one but as purely the statistical one. 
In order to {\it specify} the statistical ensemble 
in the geometrically-meaningful way, 
we prepare the following 3 dimensional hyper-surface in dS$_4$ space-time 
based on the isotropy requirement in space ($x,y,z$):\ 
$\ \bm{<<}\ 
x^2 + y^2 + z^2 = r(t)^2, 
\ (23b 
)\bm{>>}\ $ 
where $r(t)$ is the radius of S$^2$ in the 3D plane 
standing at t of the time axis. See Fig.4. $r(t)$ is an arbitrary function and will be used 
for the averaging by considering all possible forms. 

The hyper-surface is specified by $r(t)$. 
In the following path-integral, we regard the hyper-surface as a (generalized) path. 
On the path (23b), 
we obtain the {\it induced} metric $g_{ij}$ as
\ \ 
$\ \bm{<<}\ 
ds^2=g^{infl}_\mn dx^\m dx^\n = -dt^2+\e^{2\HU t}dx^idx^i
=(-\frac{1}{r^2\rdot^2}x^ix^j+\del^{ij}\e^{2\HU t})dx^idx^j
\equiv g_{ij}dx^idx^j
,
\ (24)\bm{>>}\ $
\ \ 
The constraint in (23) reduces to
\ \ 
$\ \bm{<<}\ 
\{\half(\pl_if^j+\pl_jf^i)v^j+2\pl_0f^i\}v^i=0,\ v^i\equiv\frac{dx^i}{dt}
,\ f^i=f^i(\xvec(t),t),\ 
\half v^i\vvec\cdot\pl_i\fvec+\half\{(\vvec\cdot\na)\fvec\}\cdot\vvec
+\pl_0\fvec\cdot\vvec=0
,
\ (25)\bm{>>}\ $
\ \ 
In the last formula, Lagrange derivative (which is used in the hydrodynamics) appears. 
As the geometrical quantity to define the statistical distribution, 
we take the {\it area} $A$ of the hypersurface. 
\ \ 
$\ \bm{<<}\ 
A[x^i,\xdot^i]=\int\sqrt{\det g_{ij}}~d^3\xvec=
\frac{2\sqrt{2}}{3}\int_0^\ell\e^{-3\HU t}\sqrt{\rdot^2-\e^{-2\HU t}}~dt
.
\ (26)\bm{>>}\ $
\ \ 
Hence the averaged (over the fluctuation) effective action is given by 
\ \ 
$\bm{<<}\ 
\Ga^{avg}[\Phicl;g_\mn]
=\int_{1/\La}^{1/\mu}d\rho\int_{r(0)=\rho,r(\ell)=\rho}\Dcal x^i(t)\times
\Ga[\Phicl(\xvec(\ttil),\ttil);g_\mn(\xvec(\ttil),\ttil)]\exp(-\frac{1}{2\al'} A[x^i,\xdot^i]). 
\ (26b)\bm{>>}$ 
\ \ 
where
$\m, \La$ and $\frac{1}{\al'}$ are IR cutoff, UV cutoff and the surface tension parameter
respectively. 

More generally, we can consider, instead of $A/2\al'$, 
\ \ 
$\ \bm{<<}\ 
H[x^i,\xdot^i]=\int\sqrt{\det g_{ij}}~( {1}/{2\al'}+{1}/{2\ka}\cdot {\hat R}(g_{ij})+O({\pl_i}^4)) d^3\xvec,\ 
\ (27)\bm{>>}\ $
\ \ 
where $\ka$ is the gravitational constant on the 3D hyper-surface. 
${\hat R}(g_{ij})$ is the curvature of the 3D hyper-surface. 
We regard $\al'$ and $\ka$ as model-parameters to describe the power spectrum. 

\begin{figure}[h]
    \begin{minipage}{16pc}
\includegraphics[width=16pc,height=9pc]{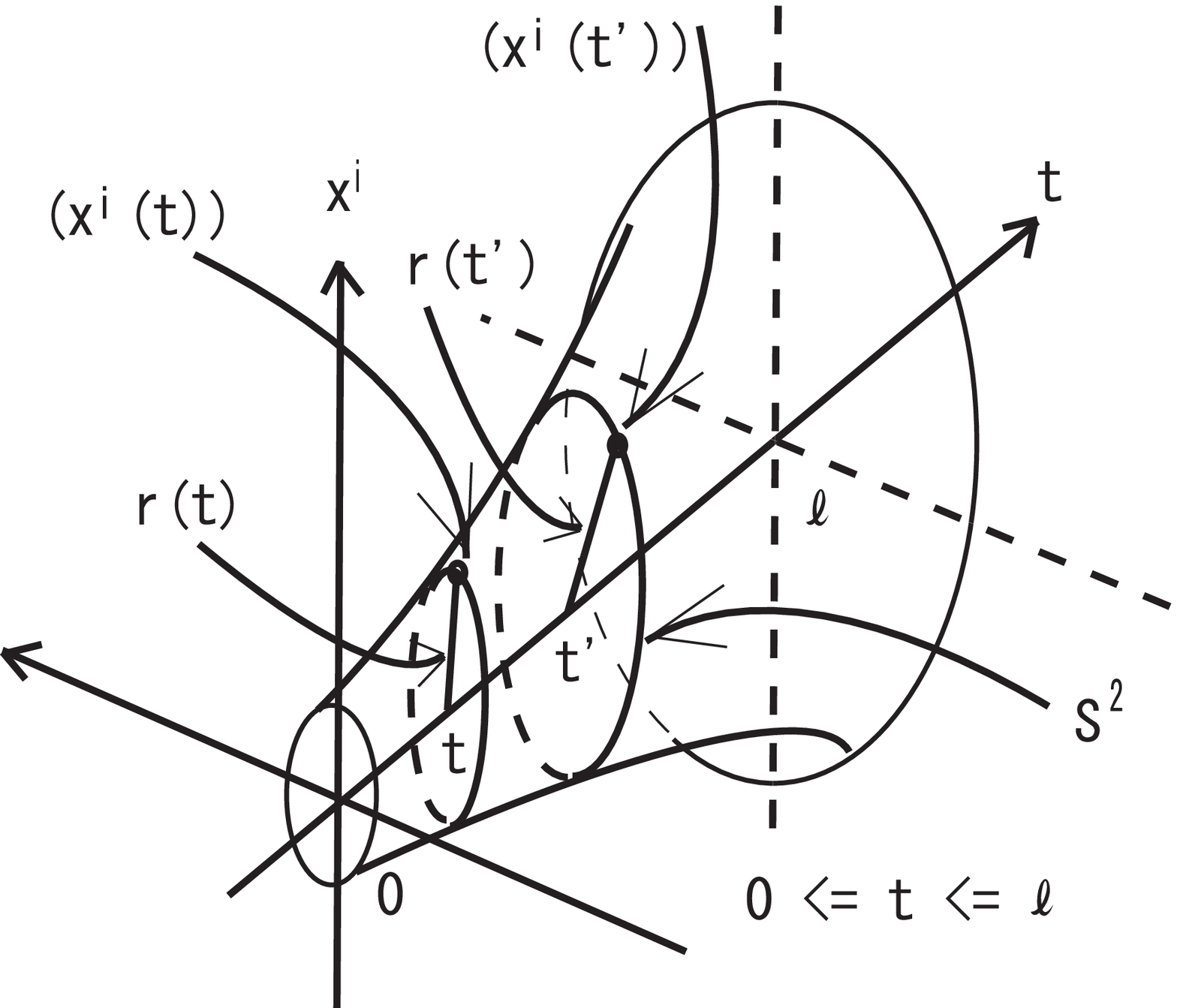}
\caption{\label{3dHSdS4}
Two points $\xvec(t), \xvec(t')$ in (28).
           }
    \end{minipage} 
\hspace{0pc}
    \begin{minipage}{16pc}
\includegraphics[width=16pc,height=9pc]{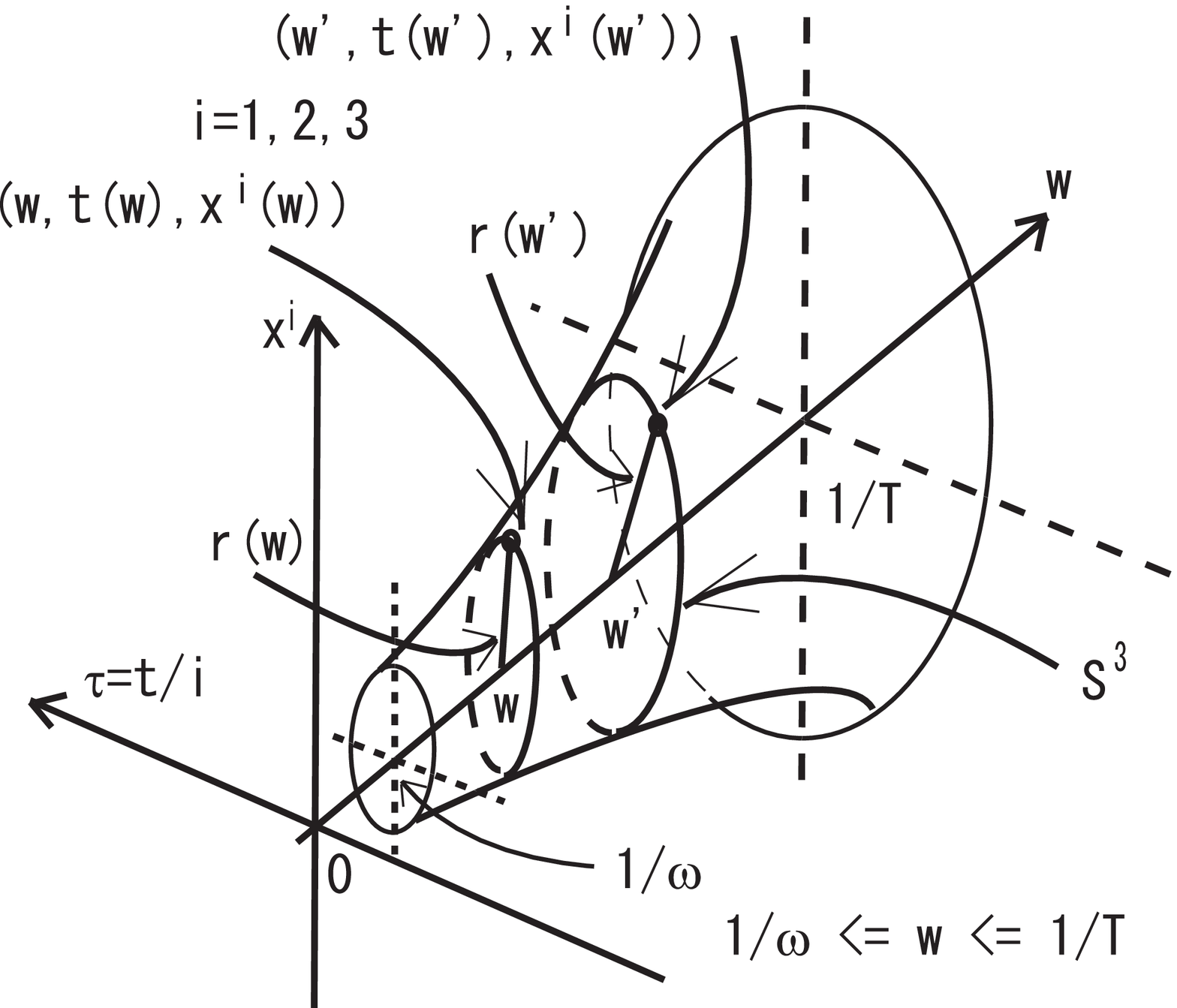}
\caption{\label{4dHSAdS5}
Two points ($t(w), \xvec(w)$), ($t(w'), \xvec(w')$) in (29). 
$t(w)=t(w')=i\tau\ ,\ \xvec\cdot\xvec+\tau^2=r^2(w)\ ,\ 
\xvec'\cdot\xvec'+\tau^2=r^2(w')$\ .
            }
    \end{minipage}
%
%
\end{figure}

The 2-point function of the spectrum is given by[1]
\nl
$\ \bm{<<}\ 
\mbox{2-Point Function}\ \ 
\del^2\Ga^{avg}/\del\Ptil_0(t)\del\Ptil_0(t'),\  
\Ptil_0(t_1)\equiv\Phi_0(\xvec(t_1),t_1), 
\ (28)\bm{>>}\ $
\ \ 
See Fig.\ref{3dHSdS4}. 


Natural extension of the above model is AdS$_5$ extra-dimensional one.\\  
$\ \bm{<<}\ 
\mbox{2-Point Function}\ \ 
\del^2\Ga^{avg}/\del\Ptil_0(w)\del\Ptil_0(w'),\  
\Ptil_0(w_1)\equiv\Phi_0(\xvec(w_1),t(w_1),w_1), t(w)=t(w')=i\tau
.
\ (29)\bm{>>}$ 
\ \ 
See Fig.\ref{4dHSAdS5}.  

\section{Discussion and Conclusion\label{conc}}
We have presented a new approach to the cosmological fluctuation[4,5,6,7]. It is regarded 
as the statistical fluctuation of space-coordinates due to the un-known "micro" dynamics. 
We adopt the (generalized) path-integral formalism to introduce the statistical ensemble. 
In the formulation, the geometric object (area $A$, (26)) is taken 
as the key quantity which determines the statistical distribution. In this new model 
some new parameters appear:\ surface tension ($1/\al'$), 
IR parameter ($\ell$), IR cut-off ($\m$), UV cut-off ($\La$), etc.. 
They are taken to explain the observational data.  \nl
\nl
{\bf\Large References}:\nl
[1]\ L.D. Faddeev and A.A. Slavnov, 
"Gauge Fields: Introduction to Quantum Theory", Benjamin/Cummings Pub., C1980, 
and references therein; \nl
[2]\ E. Verlinde, arXiv:1001.0785[hep-th] and references therein;\nl
[3]\ L. Randall and M. D. Schwartz, JHEP {\bfseries 0111:003}, 2001, arXiv:0108114[hep-th]: \nl
[4]\ S. Ichinose, 
Jour. Phys. Conf. Ser. {\bfseries 384}(2012)012028, arXiv:1205.1316; \nl
[5]\ S. Ichinose, 
Jour.Phys.:Conf.Scr.{\bfseries 222}(2010)012048, arXiv:1001.0222;\nl 
[6]\ S. Ichinose, Prog.Theor.Phys.
{\bfseries 121}(2009)727, ArXiv:0801.3064v8; \nl
[7]\ S. Ichinose, 
"Geometric Approach to Quant. Stat. Mech. and Minimal Area Principle", 
arXiv:1004.2573.     

\end{document}